\documentclass[twocolumn,showpacs,preprintnumbers,amsmath,amssymb]{revtex4}

\usepackage{graphicx}% Include figure files
\usepackage{dcolumn}% Align table columns on decimal point
\usepackage{bm}% bold math

\begin{document}

\title{Fluidization of granular media wetted by liquid $^4$He}% Force line breaks with \\

\author{K. Huang}
\email{kai.huang@ds.mpg.de}
\author{M. Sohaili}
\author{M. Schr\"oter }
\author{S. Herminghaus}
\email{stephan.herminghaus@ds.mpg.de} \affiliation{Max Planck
Institute for Dynamics and Self-Organization, Bunsenstr.10, 37073
G\"ottingen, Germany}

\date{\today}

\begin{abstract}
We explore experimentally the fluidization of vertically agitated
PMMA spheres wetted by liquid $^4$He. By controlling the
temperature around the $\lambda$ point we change the properties of
the wetting liquid from a normal fluid (helium I) to a superfluid
(helium II). For wetting by helium I, the critical acceleration
for fluidization ($\Gamma_c$) shows a steep increase close to the
saturation of the vapor pressure in the sample cell. For helium II
wetting, $\Gamma_c$ starts to increase at about 75\% saturation,
indicating that capillary bridges are enhanced by the superflow of
unsaturated helium film. Above saturation, $\Gamma_c$ enters a
plateau regime where the capillary force between particles is
independent of the bridge volume. The plateau value is found to
vary with temperature and shows a peak at 2.1\,K, which we
attribute to the influence of the specific heat of liquid helium.

\end{abstract}

\pacs{45.70.-n, 68.08.Bc, 67.25.dm}% PACS
\maketitle

It is a well known experience that the addition of a certain
amount of wetting liquid to a pile of sand increases its
mechanical stability dramatically \cite{Duran00, Nagel92,
Hornbach97, Schiffer05, Nowak05}, leading to a material stiff
enough for sculpting sand castles. The increased mechanical
stability of wet granulates is due to the formation of liquid
bridges between adjacent grains which exert attractive forces by
virtue of their surface tension \cite{Herminghaus05, Bocquet98,
Bocquet02, Levine98, Pietsch68}. As it has recently been shown,
the presence of liquid changes as well the acoustic properties of
the granulate \cite{Brunet08} Wet granular media exist in many
chemical, pharmaceutical or food production processes where the
question how to handle them appropriately is of great economic
significance \cite{Rumpf62, Iveson01, Groeger03}. Moreover, wet
granular media are also also model systems to study phase
transitions far from equilibrium \cite{Fingerle08}. A detailed
understanding of the interaction between the liquid and the grains
is therefore of major importance.

In both basic research and many industrial processes, vertical
vibration is a widely used fluidization scheme \cite{Gutman68}.
For wet granulates, an extra force must be exerted in wet grains
to overcome the cohesive capillary forces, in contrast to the
fluidization of a dry granular pile
\cite{Shapori99,Andreas06,Mujica98,Urbach05}. These forces
increase the critical shaking acceleration $\Gamma_c$ needed for
fluidization, which makes $\Gamma_c$ a good parameter to study the
influence of a wetting liquid \cite{ Scheel04,Scheel07,
Fournier05}.

Here we measure $\Gamma_c$ for PMMA spheres wetted by liquid
helium. Helium wets most substrates perfectly \cite{Cheng91,
Wyatt95, Bigelow92} so that a zero contact angle can be assumed.
When its temperature is below the $\lambda$ point (of 2.17 Kelvin
for bulk helium), liquid helium will undergo a phase transition
into a superfluid (helium II) where many interesting phenomena
such as the `fountain effect' arise, owing to its two-fluid
properties \cite{WilkBook, Allen38, London39, Meyer52, Meyer70,
Atkins59}. In this paper we study how the difference between a
superfluid and a normal fluid changes the mechanical properties of
a granular medium wetted by this liquid.

%\textit{Experimental Setup}
\begin{figure}
\includegraphics[width = 0.45\textwidth]{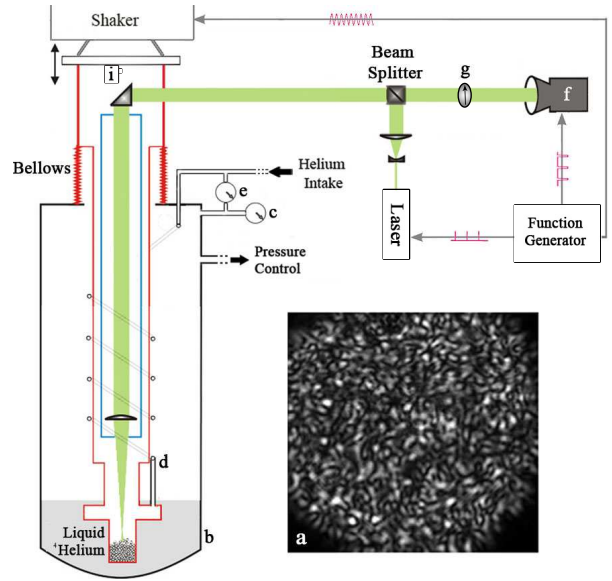}% Here is how to import EPS art
\caption{\label{setup}(Color Online) Sketch of the experimental
setup. (a) is a typical top view of the sample. The temperature in
the cryostat (b) is adjusted by controlling the pressure (measured
by pressure gauge (c)) above the liquid helium. Helium gas is
added into the sample chamber through a capillary (d). Gauge (e)
measures the pressure difference between cell and cryostat. The
sample is illuminated by laser pulses, and images are taken by a
CCD camera (f) with a polarizer (g) in front. The sample cell can
be vibrated vertically by an electrodynamic shaker (LDS V555)
which is mounted upside down above the cryostat. The strength of
vibration is measured by an accelerometer (PCB Piezotronics
353B33) (i).}
\end{figure}

A sketch of the experimental setup is shown in Fig.~\ref{setup}.
The granular sample consists of 0.6\,g PolyMethylMethAcrylate
(PMMA) spheres (Bangs Labs) with an average diameter of $d \approx
10\,\mu$m and $15\%$ width of the size distribution. This prevents
the formation of a crystalline packing which would result in
unwanted side effects. The sample is fluidized by sinusoidal
vertical vibrations with a driving frequency $f$ = 110\,Hz and a
non-dimensional acceleration $\Gamma=4\pi^2f^2A/g$, where $g$ is
the gravitational acceleration and $A$ is the shaking amplitude.
The sample is contained in a cylindrical cell  made of 99.95\%
oxygen-free copper which ensures good thermal contact with the
surrounding helium bath. The temperature in the cryostat is
controlled by adjusting the pressure above the liquid helium.

The amount of helium in the sample cell is controlled by adding
room temperature helium gas, which was passed through a cold trap
for purification. Well defined amount of helium gas was admitted
to the cell using a gauged cylinder volume. All measurements are
taken after the pressure in the cell becomes stable.

The sample was illuminated with laser pulses with 532\,nm
wavelength and a repetition rate of 22\,Hz, phase locked to the
vibration of the sample. The speckle pattern from the
back-scattered light is captured with a Charge Coupled Device
(CCD) camera (Hamamatsu C9300) with a quantum efficiency 58\% at
this wavelength. A polarizer in front of the camera suppresses
directly reflected light. The camera and laser are synchronized so
that the images are taken at a fixed phase of every fifth
vibration cycle. The power injected by the laser pulses is on the
order of $10^{-6}$\,W. This is at least one order of magnitude
less than the energy injected by vibrations which we estimate to
be $\approx 2.6\times 10^{-5}$\,W from the inelastic collisions
between the sample and the bottom plate at $\Gamma=2$.

To create a reproducible initial packing, we first fully fluidize
the sample by shaking it for a few seconds with a $\Gamma$ of
6$\pm$1. Then we ramp $\Gamma$ down  to below 1 during
approximately one minute. The critical acceleration for
fluidization $\Gamma_c$ is then measured by slowly increasing and
decreasing $\Gamma$. $\Gamma_c$ differs by maximally 15\% for
fluidization and solidification \cite{Fournier05}, the values
reported here are averages, between both values. The transition
between solid and fluid states of the sample is determined by
visual inspection of the variation of speckle pattern in real
time. As soon as the sample fluidize, the speckle pattern no
longer stays stable and starts to vary with time. This method was
found to agree with measurements based on the correlation of
subsequent images.

%\emph{Results and discussion}
\begin{figure}
\includegraphics[width=0.4\textwidth]{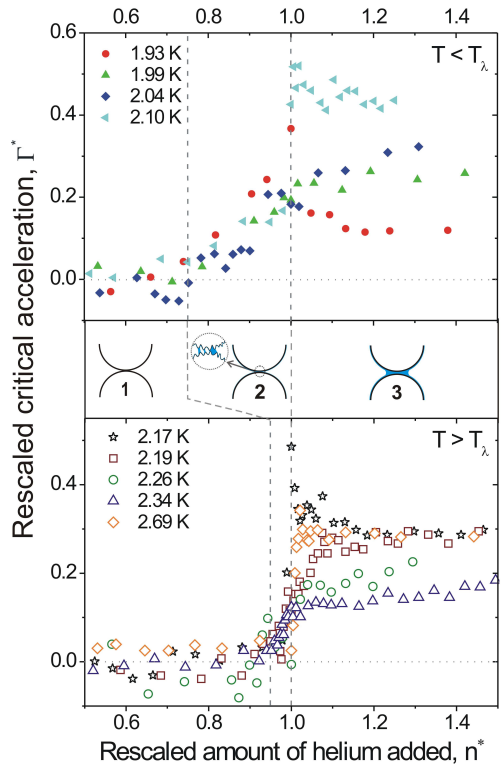}
\caption{\label{vardP}(Color Online) Scaled critical acceleration
for fluidization ${\Gamma^*}$ as a function of the amount of
helium added $n^*$ for superfluid (top) and normal fluid (bottom)
wetting. $\Gamma^*$ = 0 corresponds to the behavior of dry grains,
$n^*$ = 1 to the transition from an un-saturated to a saturated
helium film. Three different regimes can be distinguished: dry
(1),  asperity(roughness) (2) and complete wetting (3). Sketches
of the three regimes on the grain size scale are shown in the
middle. The dashed lines separating the regimes are guides to the
eye. Each data point is an average of three measurements; the
standard error is within 0.05.}
\end{figure}

Fig.~\ref{vardP} shows the dependence of the critical acceleration
for fluidization on the amount of helium gas added at different
temperatures around the $\lambda$ point. For clearer display we
use a scaled critical acceleration for fluidization
$\Gamma^*=(\Gamma_c-\Gamma_{dry})/\Gamma_{dry}$. $\Gamma_{dry}$,
the fluidization acceleration for a dry sample, is an average of
$\Gamma_c$ before it increases due to wetting.

Moreover we scale the amount of helium added, $n$, by the amount
of helium gas needed for the pressure in the cell to reach
saturation $n_{sat}$ by defining $n^*=n/n_{sat}$. For $n^* < 1$,
$n^*$ can be treated as the fractional saturation of helium,
$P_c/P_0$, in the cell, where $P_c$ is the pressure in the cell
and $P_0$ the saturated vapor pressure, because most of the helium
added stays in the vapor phase. For $n^* > 1$, $n^* -1$ grows
linearly with liquid content W, by which we denote the ratio of
the volume of the wetting liquid and the total volume occupied by
the sample. The data with largest $n^*$ shown below correspond to
a liquid content varying from 10 \% to 38 \% at different
temperature.

%This assumption is justified by the linearity of a plot of
%$P_c/P_0$ versus $\Delta n^*$: less than $3\%$ of the added helium
%condenses on the particles.

Fig.~\ref{vardP} shows that the behavior of ${\Gamma^*}$ can be
divided in three distinct regimes. In regime $1$  the sample
behaves the same as a dry granular medium; the scaled critical
acceleration stays around zero. Adding helium gas in this regime
leads to the increase of pressure in the cell and the formation of
the first atomic layer of helium on the particles. This layer is,
however, not mobile enough to form liquid bridges \cite{WilkBook}.

Regime $2$ corresponds to the asperity wetting regime, where
${\Gamma^*}$ increases monotonically with the amount of added
helium. In this regime the helium film condensed on the particle
surface is thick enough to form small liquid bridges between the
asperities of adjacent particles. With the increase of helium
adsorbed, the number of small capillary bridges at asperity level
increases, which leads to higher cohesive force between adjacent
particles.

In regime $3$, the amount of adsorbed helium is enough to fill the
roughness on the grains, such that they appear as completely wet,
perfect spheres to all further added liquid. $\Gamma^*$ shown in
Fig.~\ref{vardP}, within experimental scattering, stays constant,
in agreement with earlier experiments with other liquids at room
temperature \cite{Scheel04}. This independence from $n^*$ comes
from the fact that the capillary force is dominated by the
curvature of the spheres instead of the volume of the liquid
bridges. With the increase of $n^*$ capillary bridges will
coalesce and form bigger liquid clusters, but ${\Gamma^*}$ will
not change due to the constant Laplace pressure imposed by the
packing geometry\cite{Scheel07}.

It is regime 2 where the difference between superfluid and normal
fluid wetting is significant. For superfluid wetting the sample
enters regime 2 already at a fractional saturation of about 0.75,
which is far below the 0.95 observed for normal fluid wetting. We
interpret the increase of ${\Gamma^*}$ to be due to the formation
of bridges between asperities of neighboring spheres by the
condensed unsaturated helium film. The amount of helium adsorbed
can be described by the Frenkel-Hasley-Hill equation
\cite{Bowers53,WettingReview}; it is proportional to
$(-ln(P_{c}/P_{0}))^{-1/3}$ and therefore as shown above to $(-ln
(n^*))^{-1/3}$.

\begin{figure}
\includegraphics[width=0.45\textwidth]{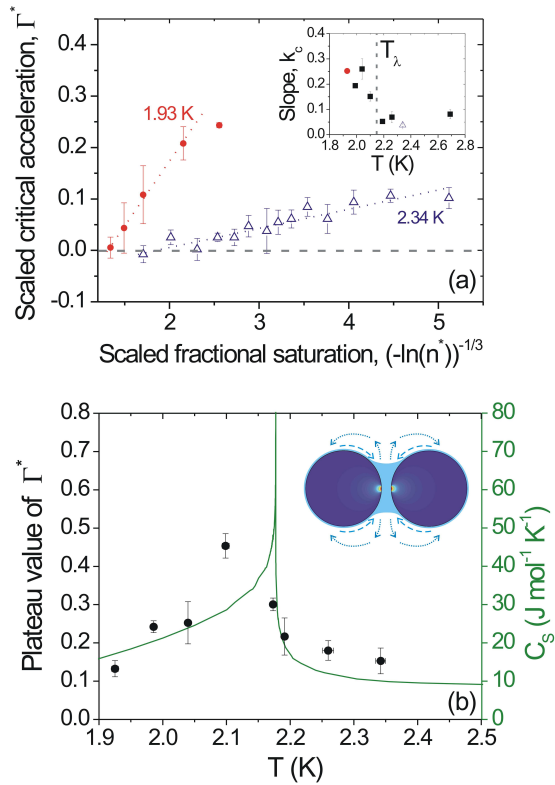}
\caption{\label{varT}(Color Online) (a) In regime 2 ${\Gamma^*}$
varies linearly with the scaled saturation $(-ln (n^*))^{-1/3}$.
Dotted lines are fits with Eqn.~\ref{eqn:two}. The inset of (a)
shows the temperature dependence of the fit parameter $k_c$. (b)
The temperature dependence of plateau value of ${\Gamma^*}$ in
regime 3 (black dots). The dark green line is the specific heat
$C_s$ of bulk helium taken from \cite{Donnelly77}. The inset of
(b) shows a sketch of two completely wetted particles. The color
code illustrates the temperature gradients generated by energy
dissipation. Blue dash arrows indicate the flow of superfluid in
the helium film driven by `fountain effect' and blue dotted arrows
depict evaporation of helium from the surface of meniscus.}
\end{figure}

Fig.~\ref{varT}(a) shows that $\Gamma^*$ increases linearly in regime 2
 with the unsaturated helium
film thickness; both above and below the $\lambda$ point. This can
be understood in the following way. To fluidize wet granular
media, the driving force has to overcome the capillary forces
between the grains or between container walls and the
particles. The gravitational force can be neglected here
because it is two orders of magnitude smaller than the capillary
force. In the asperity regime $2$, the capillary force $f_b$ is
given by \cite{Herminghaus05}:
\begin{equation}
f_b= \frac{R^2}{2\pi \delta^2} \, \tilde{V}\, F_b,
\label{cohesiveforce}
\end{equation}
where
\begin{equation}
F_b=2\pi R \gamma cos(\theta),
\label{eqn:one}
\end{equation}
is the capillary force in the complete wetting regime, R is the
radius of the particle, $\tilde{V}$ is the bridge volume, which
depends linearly on the amount of helium adsorbed, $\delta$ is the
amplitude of the roughness of the particles and $\theta$ is the
contact angle. The linear growth of the cohesive force with the
bridge volume in this regime explains the linear dependence of
$\Gamma^*$ on the amount of helium adsorbed
(eq.~(\ref{cohesiveforce})). Therefore we fit in
Fig.~\ref{varT}(a) the values of $\Gamma^*$ with
\begin{equation}
\Gamma^*=k_c (-ln (n^*))^{-1/3} + \beta, \label{eqn:two}
\end{equation}
with the slope coefficient $k_c$ and $\beta$ as fitting
parameters.

The temperature dependence of $k_c$ shown in the inset of
Fig.~\ref{varT}(a) depicts the enhancement of the cohesive force
by adsorbed superfluid film below the $\lambda$ point. This is
readily explained from the strongly different transport mechanisms
in the superfluid state. First, the superflow enables the forming
capillary bridge to acquire more liquid from its surrounding
during bridge formation. Second, the impact of the spheres
radiates quantum excitations into the superfluid at the point of
contact, dragging extra superfluid towards the contact region by
osmosis (fountain effect) \cite{Allen38, London39}.  This is in
contrast to normal fluid wetting, where only liquid very close to
the contact point is sucked into the bridge by the negative
Laplace pressure.

As it is clearly seen in Fig.~\ref{vardP}, the critical
acceleration  shows a plateau in regime 3 both above and below
$T_{\lambda}$, as it is observed as well with standard liquids
\cite{Scheel04,Fournier05,Herminghaus05}. The constancy of
$\Gamma^{\ast}$ reflects the weak dependence of the capillary
force upon liquid volume for fully developed capillary bridges
\cite{Herminghaus05,Scheel07}. However, since the surface tension
depends only weakly on temperature close to $T_{\lambda}$, we
would expect to observe roughly the same plateau value for all
temperatures, which is clearly not the case.

Fig.~\ref{varT}(b) shows the temperature dependence of the plateau
value of $\Gamma^*$ in regime 3. It shows a peak at about 2.1\,K,
close to the superfluid transition. This can be qualitatively
understood as a combination of two effects. In the normal fluid
regime, capillary bridges will acquire their full volume only
close to $T_{\lambda}$, where the specific heat of the liquid
(shown for comparison) is large and prevents strong heating of the
bridge from dissipated energy. Farther away from $T_{\lambda}$,
the bridges heat up and evaporate back into the asperity regime.
In the superfluid regime, temperature is effectively equalized by
the superflow. The heat intake due to bridge rupture and grain
impact thus gives rise to a strong superflow towards the contact
points. However, it is well known that the presence of a superflow
in an adsorbed liquid film leads to a strongly increased contact
angle $\theta$ \cite{Herminghaus98}. This is due to the
Kontorovich pressure term \cite{Kontorovich56} adding to the
disjoining pressure, and leads to dynamical incomplete wetting in
liquid helium, as observed experimentally \cite{Rolley02}. This
effect is weak close to $T_{\lambda}$, but increases further into
the superfluid regime. As a consequence, the capillary force (and
thus the plateau value of $\Gamma^{\ast}$) is reduced according to
eq.~(\ref{eqn:one}).

Need to mention that no clear signature of viscous effect can be
found. This can be understood by calculating the ratio between the
energy dissipation by viscosity and by rupture of capillary
bridges \cite{Herminghaus05}. At temperature 2.1\,K, it yields
0.0046, indicating that the system is still in the capillary
region where viscosity can be ignored.

To conclude, we demonstrate with liquid helium, a liquid that has
a surface tension only 1/200 of pure water, that the increase of
mechanical stability of granular materials by wetting is
prominent. Beyond that, our main findings can still be explained
on the basis of a simple capillary model \cite{Herminghaus05} by
taking superfluid properties of liquid helium, such as `fountain
effect', into account.

%Besides, we also show that $\Gamma_c$ can be a nice probe to study
%capillary condensation of superfluid helium in a granular system.

Inspiring discussions with Mario Scheel, Axel Fingerle, Martin
Brinkmann, J\"urgen Vollmer, and Isaac Goldhirsch are gratefully
acknowledged. We thank Udo Krafft and G\"unter von Roden for their
indispensable technical support.

\end{document}